\documentclass[%
superscriptaddress,
frontmatterverbose,  
preprint,
showpacs,
preprintnumbers,
amsmath,
amssymb, 
aps, 
12pt,
pre,
floatfix 
]{revtex4-1}

\usepackage{amsfonts}       
\usepackage{amsmath}       
\usepackage{amssymb}
\usepackage{nicefrac}
\usepackage{float}
\usepackage[lofdepth,lotdepth]{subfig}
\usepackage{graphicx}
\usepackage{dcolumn}
\usepackage{bm}
\usepackage{hyperref}
\usepackage{xcolor}  
\usepackage{array}
\usepackage[export]{adjustbox}
\usepackage{multirow}
\usepackage[figure,table]{totalcount}
\usepackage{verbatim}
\usepackage{ragged2e}
\usepackage[font=small,labelfont=bf,
   justification=justified,
   format=plain]{caption}

\newcommand{\avg}[1]{\left\langle#1\right\rangle} 

\begin{document}

\title{Mechanisms of self-organized quasicriticality in neuronal networks models}

\author{Osame Kinouchi}

\email{okinouchi@gmail.com}
\thanks{corresponding author}

\affiliation{Universidade de S\~ao Paulo, FFCLRP, 
Departamento de F\'isica, Ribeir\~ao Preto, SP, 14040-901, Brazil}

\author{Renata Pazzini}
\email{rpazzini@usp.br}

\affiliation{Universidade de S\~ao Paulo, FFCLRP, 
Departamento de F\'isica, Ribeir\~ao Preto, SP, 14040-901, Brazil}

\author{Mauro Copelli}

\email{mcopelli@gmail.com}

\affiliation{Departamento de Física, Universidade Federal de Pernambuco, Recife, Brazil, Brazil}

\begin{abstract}
The critical brain hypothesis states that
there are information processing advantages for
neuronal networks working close to the critical region of
a phase transition. 
If this is true, we must ask  how the
networks achieve and maintain this critical state. 
Here we review several proposed
biological mechanisms that turn the critical region 
into an attractor of a dynamics in  network
parameters like synapses, neuronal gains and firing thresholds. 
Since  neuronal networks (biological and models) 
are nonconservative but dissipative, we expect 
not exact criticality but self-organized quasicriticality 
(SOqC), where the system hovers around the critical point.
\end{abstract}


\maketitle

\section{Introduction}

Thirty-three years after the initial formulation of the 
self-organized criticality 
(SOC) concept~\citep{Bak1987} (and thirty-seven years after of the self-organizing
extremal invasion percolation model~\cite{Wilkinson1983}), one of the
most active areas that employ these ideas is theoretical neuroscience.
However, neuronal networks, similarly to earthquakes and forest
fires, are nonconservative systems, in contrast to canonical SOC systems like sandpile models~\citep{Jensen1998,Pruessner2012}. 
To model such systems one uses  nonconservative
networks of elements represented 
by cellular automata, discrete time
maps or differential equations. Such models have
distinct features from conservative systems. 
A large fraction of them, in particular neuronal networks, have been described as displaying self-organized quasi-criticality (SOqC)~\cite{Bonachela2009,Bonachela2010,Buendia2020b} or weak criticality~\cite{Palmieri2018,Palmieri2020}, which is the subject of this review.

The first person that made an analogy
between brain activity and a critical 
branching process probably was Alan Turing,
in his memorable paper \emph{Computing machinery and intelligence}~\citep{Turing1950}. Decades later,
the idea that SOC models could be important to describe the 
activity of neuronal networks was in the air as early as 1995~\citep{Usher1995,Corral1995,Bottani1995,Chen1995,Herz1995,Middleton1995}, eight years before the fundamental 2003 experimental paper of Beggs and
Plenz~\citep{Beggs2003} reporting 
neuronal avalanches.
This occurred because several authors, working with
models for earthquakes and pulse-coupled threshold
elements, noticed the formal analogy
between such systems and networks of integrate-and-fire neurons.
Critical learning was also conjectured by Chialvo and
Bak~\citep{Stassinopoulos1995,Chialvo1999,Bak2001}.
However, in the absence of experimental support, these
works, although prescient, were basically 
theoretical conjectures. A historical question would be to
determine in what extent this early literature
motivated Beggs and Plenz to perform their 
experiments.

Since 2003, however, the study of criticality 
in neuronal
networks developed itself as a research paradigm, with
a large literature,
diverse experimental approaches and several 
problems addressed theoretically and 
computationally (some reviews include Refs.~\citep{Chialvo2010,Hesse2014,Plenz2014,
Cocchi2017,Munoz2018,Wilting2019,Buendia2020b}).
One of the main results is that  information
processing seems to be optimized at a second order
absorbing phase
transition~\citep{Haldeman2005,Kinouchi2006,
Copelli2007,Wu2007,Assis2008,
Beggs2008,Ribeiro2008,Shew2009,Larremore2011,
Shew2011,Shew2013,Mosqueiro2013,Wang2017,Zierenberg2020,
Galera2020}. 
This transition occurs  between no activity 
(the absorbing phase) and
nonzero steady state activity (the active phase). 
Such transition is familiar from the SOC literature
and pertains to the Directed Percolation (DP) or the
Conservative-DP (C-DP or Manna) universality 
classes~\citep{Buendia2020b,Galera2020,
Dickman1998,Munoz1999,Dickman2000}. 

An important question is how neuronal 
networks self-organize toward the critical region. 
The question arises because, like earthquake and forest-fire models, neuronal networks are not conservative systems, which means that in principle they cannot be exactly critical~\citep{Buendia2020,Dickman2000,Bonachela2009,Bonachela2010}.
In these networks, we can vary control parameters like the strength of synapses and obtain subcritical, critical and supercritical behavior. 
The critical point is therefore achieved only by fine-tuning. 

Over time, several authors proposed different
biological  mechanisms that could eliminate the
fine-tuning and make the 
critical region a self-organized attractor.
The obtained criticality is not perfect, but it is
sufficient to account for the experimental data.
Also, the mechanisms (mainly based on 
dynamic synapses but also on dynamic neuronal 
gains and adaptive firing thresholds)
are biologically plausible and should be viewed as a
research topic \emph{per se}.

The literature about these homeostatic mechanisms 
is vast and we do not intend to
present an exhaustive review. However, we discuss here
some  prototypical  mechanisms and try to connect them to
self-organized quasicriticality (SOqC), a concept 
developed to account for non-conservative systems that hover around but
do not exactly sit on the critical
point~\citep{Buendia2020b,Bonachela2009,Bonachela2010}.

For a better comparison between the models, we will not rely on the original notation of the reviewed articles, but will try to use a universal notation instead. For example,  the synaptic strength
between a presynaptic neuron $j$ and a postsynaptic neuron
$i$ will be always denoted by $W_{ij}$
(notice the convention in the order of the indexes), 
the membrane potential is $V_i$, the
binary firing state is $s_i \in \{0,1\}$,
the gain of the firing function is $\Gamma_i$, 
the firing threshold is $\theta_i$ and so on.
To prevent an excess of index subscripts 
as is usual in dynamical systems, like $W_{ij,t}$, we  
use the convention $W_{ij}(t)$ for continuous time
and $W_{ij}[t]$ for discrete time.

Lastly, before we begin, a few words about the fine-tuning problem. 
Even perfect SOC systems are in a sense fine tuned: they 
must be conservative and
require infinite separation of time 
scales with driving rate $
1/\tau \rightarrow 0^+$ and dissipation rate $u \rightarrow 0^+$ with  $1/(\tau u) \rightarrow 0$~\cite{Buendia2020b,Dickman1998,Dickman2000,Jensen1998,Pruessner2012}. 
For homeostatic systems, we turn a control parameter like the coupling $W$ in a time-dependent slow variable $W[t] = \avg{W_{ij}[t]}$ by imposing a local dynamics in the individual $W_{ij}$.
This  dynamics could depend on new parameters (here called hyperparameters) which need some  tuning (in some cases this tuning can be very coarse in the large $\tau$ case). 
Have we exchanged the fine tuning on $W$
by several tuning operations on the homeostatic 
hyperparameters?
Not exactly, as nicely discussed by
Hernandez-Urbina and Herrmann~\citep{Hernandez2017}:

\begin{quote}
\begin{center}{\bf To Tune or Not to Tune}\end{center}
 In this paper, we have shown how systems self-organize into a critical state through [homeostasis]. Thus, we became relieved from the task of fine-tuning the control parameter $W$, but instead we acquire a new task: that of estimating the appropriate values for parameters $A, B, C$, and $D$. Is there no way to be relieved from tuning any parameter in the system?

The issue of tuning or not tuning depends mainly on what we understand by control parameter. (...) a control parameter can be thought of a knob or dial that when turned the system exhibits some quantifiable change. We say that the system self-organizes if nobody turns that knob but the system itself. In order to achieve this, the elements comprising the system require a feedback mechanism to be able to change their inner dynamics in response to their surroundings.
(...) The latter does not require an external entity to turn the dial for the system to exhibit critical dynamics. However, its internal dynamics are configured in a particular way in order to allow feedback mechanisms at the level of individual elements. 

Did we fine-tune their configuration? Yes. Otherwise, we would have not achieved what was desired, as nothing comes out of nothing. Did we change control parameter from $W$ to $A, B, C,$ and $D$? No, the control parameter is still intact, and now it is “in the hands” of the system.
(...)
Lastly and most importantly, the new configuration stresses the difference between global and local mechanisms. The control parameter $W$ (the dial) is an external quantity that observes and governs the global (i.e., the collective), whereas [homeostasis] provides the system with local mechanisms that have an effect over the collective. This is the main feature of a complex system.
\end{quote}

\section{Plastic synapses}

\begin{figure}[ht]
\begin{center}
\includegraphics[width=0.9\textwidth]{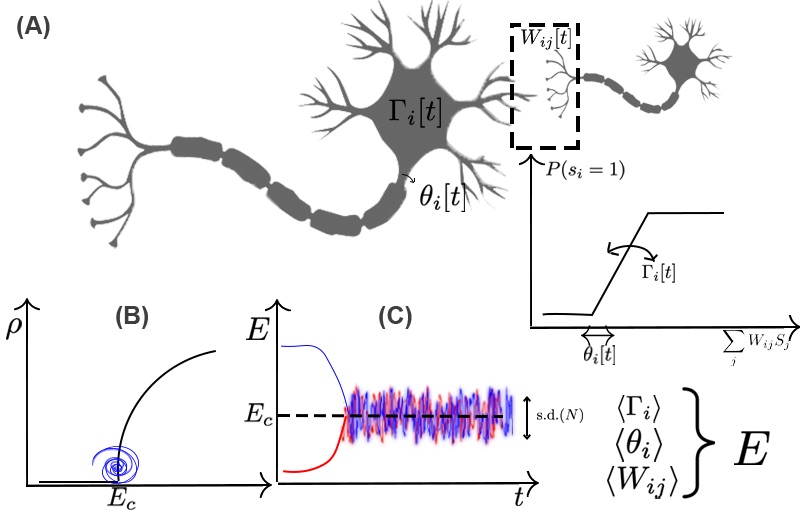}
\end{center}
\caption{
Example of homeostatic mechanisms in
a stochastic neuron with firing probability $P(s_i=1)$ \textbf{(A)} Scheme of the loci for homeostatic mechanisms: synapses $W_{ij}$, neuronal gain $\Gamma_i$ and firing threshold $\theta_i$. Inset: Firing probability
with homeostatic variables;
\textbf{(B)} Bifurcation diagram for the activity $\rho$ as a function of a generic control parameter $E$. The critical point is $E_c$ but the homeostatic fixed point (a focus) is slightly supercritical; \textbf(C) Self-organization of the
generic "control" parameter $E(t)$, 
where the standard
deviation of the stochastic oscillations around the
fixed point depends on system size as $s.d.\propto N^{-a}$.
}\label{fig1}
\end{figure}

Consider an absorbing state second order phase transition
where the activity is $\rho = 0$ below a critical point $E_c$ and
\begin{equation}\label{DP}
\rho \simeq C \left(\frac{E-E_c}{E_c}\right)^\beta\:,
\end{equation}
for $E \gtrsim E_c$, where $E$ is a generic
control parameter, see Fig.~\ref{fig1}A and B.
For topologies such as random and complete graphs, one typically obtains $\beta = 1$, which is consistent with a transition in the Mean-Field Directed Percolation (DP) class (or perhaps, the Compact-DP (Manna) class usual in SOC models, which has the same mean-field exponents but different ones below the upper critical dimension, see~\citep{Jensen1998,Lubeck2004,Buendia2020b,Galera2020}).

The basic idea underlying most of the proposed mechanism for homeostatic self-organization is to define a slow dynamics in the individual links $E_i(t)$ $(i = 1,\ldots,N)$ such that, if the network is in the subcritical state, their average value $E(t)=\avg{E_i(t)} $ grows toward $E_c$ but, if the network is in the supercritical state, $E(t)$ decreases toward $E_c$, see 
Fig.~\ref{fig1}C. 
Ideally, these mechanisms should be local, that is, they should not have access to global network information such as the density of active sites $\rho$ (the order parameter) but rather only to the local firing of the neurons connected by $E_i$. 
In the following, we give several examples from the literature.

\subsection{Short-Term Synaptic Plasticity} 

Markram and Tsodyks~\citep{Markram1996,Tsodyks1998}
proposed a short-term 
synaptic model that inpired several authors
in the area of self-organization to criticality.
The Markram-Tsodyks (MT) dynamics is:
\begin{eqnarray}
\frac{dJ_{ij}(t)}{dt} & =&  \frac{1}{\tau}
\left[\frac{A}{u(t)} - J_{ij}(t) \right] - u(t) J_{ij}(t)
\delta(t-\hat{t}_j) \:, \label{LHG} \\
\frac{du(t)}{dt} & = &  \frac{1}{\tau_u} 
\left[ U - u(t) \right]  + U\left[ 1-u(t)\right] \delta(t-\hat{t}_j) 
    \label{MT} \:,
\end{eqnarray}
where $J_{ij}$ is the available neurotransmitter
resources, $u$ is the fraction used after
the presynaptic firing at time $\hat{t}_j$ 
(so that the effective synaptic efficacy is $W_{ij}(t)
=u(t) J_{ij}(t)$), $A$ and $U$ are baseline constants (hyperparameters), and $\tau$ and $\tau_u$ are recovery time constants.

In an influential paper, Levina, Herrmann and Geisel (LHG)~\citep{Levina2007} proposed to use depressing-recovering synapses. 
In their model, we have leaky-integrate-and-fire (LIF) neurons in a complete-graph topology.
As a self-organizing mechanism, they used a simplified version of the MT dynamics with constant $u$, that is, only Eq.~\eqref{LHG}.
They studied the system varying $A$ and found
that although we need some tuning in the hyperparameter
$A$,  any initial distribution of synapses $P(W_{ij}(t=0))$
converges to a stationary distribution $P^*(W_{ij})$
with $\avg{W^*_{ij}} \approx W_c$.
We will refer to Eq.~\eqref{LHG} with constant $u$ as the
LHG dynamics. These authors found quasicriticality 
for $1.7<A<2.3, u\in]0,1]$ and $\tau \propto N$.
Levina \emph{et al.} also studied synapses with the  full MT model in~\citep{Levina2006,Levina2009}.

Bonachela \emph{et al.}~\citep{Bonachela2010} studied in depth the LHG model and found that, like forest-fire models, it is an instance of SOqC. The system presents the characteristic hovering around the critical point in the form of stochastic sawtooth  oscillations in the $W(t)$ that do not disappear in the thermodynamic limit.
Using the same model, Wang and Zhou~\citep{Wang2012} showed that the LHG dynamics also works in hierarquical modular networks, with an apparent improvement in SOqC robustness in this topology.

Note that the LHG dynamics can be written in terms of the synaptic efficacy $W_{ij}= u J_{ij}$ by multiplying Eq.~\eqref{LHG} by $u$, leading to:
\begin{equation}
\frac{dW_{ij}(t)}{dt}  = \frac{1}{\tau}
\left[A - W_{ij}(t) \right] - u W_{ij}(t)
\delta(t-\hat{t}_j) \; . \label{LHG2}
\end{equation} 

Brochini \emph{et al.}~\citep{Brochini2016}  studied a complete graph of stochastic discrete time LIFs~\citep{Gerstner1992,Galves2013} and proposed a discrete time LHG dynamics:
\begin{equation}\label{discreteLHG}
W_{ij}[t+1] = W_{ij}[t] +\frac{1}{\tau}\left(A-W_{ij}[t]
\right) - u W_{ij}[t] s_j[t] \:,
\end{equation}
where the firing index $s_j[t]\in \{0,1\}$ denotes spikes.
Kinouchi~\emph{et al.}~\citep{Kinouchi2019}, in the
same system, studied the stability of the fixed points
of the joint neuronal-LHG dynamics. 
They found that, for the average synaptic value
$W \def \avg{W_{ij}}$, the fixed point is $W^* = W_c +
\mathcal{O}((A-1)/u\tau)$, meaning that for large
$u\tau$ the systems approaches the critical point $W_c$
if $A > 1$.
However, since it is not biologically plausible to assume an infinite recovering time $\tau$, one  always obtains
a system which is slightly supercritical.
This work also showed that the fixed point is a barely
stable focus, around which the system is excited by  finite size (demographic) noise, leading to the characteristic sawtooth oscillations of SOqC. A similar scenario was already found by Grassberger for forest-fire models~\citep{Grassberger1991}.

The discrete time LHG dynamics was also studied 
for celullar automata neurons in random networks
with an average of $K$ neighbors connected by
probabilistic synapses $P_{ij}\in [0,1]$ (Costa \emph{et al.}~\citep{Costa2015}, Campos 
\emph{et al.} \citep{Campos2017} and Kinouchi \emph{et al.}~\citep{Kinouchi2019}):

\begin{equation}
    P_{ij}[t+1] = 
    P_{ij}[t] + \frac{1}{\tau}\left(
    \frac{A}{K} - P_{ij}[t]\right) -  u P_{ij}[t] s_j[t] \:, 
\end{equation}
with an upper limit $P_{\mathrm{max}}=1$.
Multiplying by $K$ and summing over $i$, we get an equation for the local branching ratio:
\begin{equation}
\sigma_j[t+1] = \sigma_j[t]+ \frac{1}{\tau}
\left( A - \sigma_j[t] \right) - 
u \sigma_j[t] s_j[t]\:.
\end{equation}
It has been found that such depressing synapses
induce correlations inside the synaptic 
matrix, affecting the global branching ratio 
$\sigma[t] = \avg{\sigma_j[t]}$,  so that 
criticality does not occur at the branching ratio $\sigma_c=1$ but rather when the
largest eigenvalue of the synaptic matrix is
$\lambda_c=1$, with $\sigma^* = K \avg{P^*_{ij}} \approx
1.1$~\citep{Campos2017}.

After examining this diverse literature, it seems that
any homeostatic dynamics of the form:
\begin{equation}\label{general}
    W_{ij}[t+1] = W_{ij}[t] + R(W_{ij}[t]) 
    - D(W_{ij},s_j[t])\:,
\end{equation}
can self-organize the networks, where $R$ and $D$ are the recovery and depressing processes, for example:
\begin{equation} \label{simple}
     W_{ij}[t+1] = W_{ij}[t] + \frac{1}{\tau} W_{ij}[t] 
    -u W_{ij}[t] s_j[t]\: .
\end{equation}
In particular, the simplest mechanism would be:
\begin{equation}
     W_{ij}[t+1] = W_{ij}[t] + \frac{1}{\tau} 
    - u s_j[t] \:,  \label{simpleLHG}
\end{equation}
a usual dynamics in SOC models~\citep{Bonachela2009,Buendia2020b}.
This means that the full LHG dynamics,
and also the full MT dynamics, is a sufficient 
but not a  necessary condition for SOqC.

The  average
$W = \avg{W_{ij}}$ for this dynamics is:
\begin{equation}
    W[t+1] = W[t] + \frac{1}{\tau} - u \rho[t] \:,
\end{equation}
where $\rho[t] = \avg{s_i[t]}$ 
is the time dependent network activity. The stationary state is $\rho^* = 1/(\tau u)$  and, if $\tau u$ is
large, this means that $\rho^* = \mathcal{O}(1/(\tau
u))\rightarrow \rho_c = 0^+$.
Also, if we use Eq.~\eqref{DP}, one gets
$W^* = W_c + \mathcal{O}(1/(\tau u))$.
The dissipative term $u$ should also be small, meaning that, if we desire absolute separation of time scales, one needs
$1/\tau \rightarrow 0^+, 
u \rightarrow 0^+, 1/(\tau u)\rightarrow 0 $,
as is usual in other SOC systems~\citep{Jensen1998,Dickman1998,
Dickman2000,Bonachela2009,Buendia2020b}.

Here, for biological plausibility, its is better 
to assume a large but finite recovery time, say
$\tau \in [100,10000]$ ms, in comparison with
$1$ ms for spikes. Also, to obtain SOqC, $u$ needs not
be small. We must have $A > 1$ because
$A < 1$ produces subcritical activity~\citep{Levina2007,Bonachela2010,Kinouchi2019}.
So, moderate $A \in [1,2]$, $u \in \:]0,1]$, and large $\tau > 1000$ seem to be the coarse tuning conditions for homeostasis.
This produces the hovering of the average value $W[t]=\avg{W_{ij}[t]}$ around the critical point $W_c$, with the characteristic sawtooth oscillations of SOqC and power-law avalanches for some decades.

We observe that the original LHG model~\citep{Levina2007,Bonachela2010} had $\tau \propto N$  to produce the infinite separation
of time scales in the large-$N$ limit. 
This, however, did not prevent the SOqC hovering stochastic oscillations in the thermodynamic limit. 
Moreover, a recovery time proportional to $N$ is a very unrealistic feature for biological synapses.
Curiously, if we use a finite $\tau u$ instead, the oscillations are damped in the thermodynamic limit because the fixed point $\rho^* = 
\mathcal{O}(1/(\tau u)),
W^* = W_c
+ \mathcal{O}(1/(\tau u))$ continues to be an attractive focus but the demographic noise vanishes.
On the other hand,  when we use  $\tau u\rightarrow \infty$, the fixed point loses its
stability and continues to be perturbed even by the $N \rightarrow \infty$ vanishing fluctuations~\citep{Kinouchi2019}.

As early as 1998, Kinouchi~\citep{Kinouchi1998}
proposed the synaptic dynamics:
\begin{equation}\label{OK}
W_{ij}[t+1] = W_{ij}[t] +\frac{1}{\tau}W_{ij}[t] - u s_j[t] \:.
\end{equation}
with small but finite $\tau$ and $u$.
The difference here from the former mechanisms is 
that, like in Eq.~\eqref{simpleLHG},
depression is not proportional to $W_{ij}$
(but recovery is).
He also discussed the several concepts of SOC at the time, and called these homeostatic system 
as self-tuned criticality
(STC), which is equivalent to a SOqC system with
finite separation of time scales.

Hsu and Beggs~\citep{Hsu2006} studied a model for the activity $A_i(t) $
of the local field potential at electrode $i$:
\begin{equation}
\label{eq:Hsu2006}
    A_i[t+1] = H_i[t] + \sum_j P_{ij}[t] s_j[t] \:,
\end{equation}
where $H_i(t)$ is a spontaneous activity used to prevent the freezing of the system in the absorbing state (this is similar to a time-dependent SOC drive term $h$). 
The probabilistic couplings is $P_{ij}\in [0,1]$. Firing rate homeostasis and critical homeostasis are achieved by increasing or
decreasing $H$ and $P$ if the firing rate is too
low or too high compared to a target firing rate
$s_0 = 1/\tau_0$:
\begin{eqnarray}
H_i[t+1] &=& \exp\left[- k_S \left(\avg{s_i[t]} - s_0
\right)  \right]  H_i[t]  \:, \\
P_{ij}[t+1] & = & \exp\left[- k_P \left(\avg{s_i[t]} - s_0
\right)   \right] P_{ij}[t] \label{Hsu}\:,
\end{eqnarray}
where $\avg{\ldots}$ represents a moving average
over a memory widow $\tau_m$. 

Hsu and Beggs found that for $k_S/k_P \approx 0.5$,
this dynamics leads to a critical 
branching ratio $\sigma = 1$. 
They also found that the target firing rate $s_0$ can be maintained by this homeostasis. 
Equation~\eqref{Hsu} reminds us of the depressing-recovering synaptic rule of Eq.~\eqref{simple}. 
Indeed, if we examine the small 
$k_P$ limit (as used by the authors), we have:
\begin{equation}
  P_{ij}[t+1]  \simeq   P_{ij}[t] + \frac{1}{\tau} P_{ij}[t] - 
  u P_{ij}[t] \avg{s_i[t]} \:,
\end{equation}
 where now $\tau = 1/(k_P s_0)$ and $u = k_P$. 
 A similar reasoning applies to the equation for $H[t]$, which could be identified with the homeostatic threshold equation~\eqref{theta} discussed in Sec.~\ref{thres}, with $H[t] = - \theta[t]$.
 
In another paper, Hsu \emph{et al.}~\citep{Hsu2007}
extended the model to include distance-dependent connectivity and Hebbian learning~\citep{Hsu2007}. Changing the homeostasis
equations to our standard notation, we have:
\begin{eqnarray}
 \frac{d H_i(t)}{dt}  & =& \frac{1}{\tau_S} (1-\eta_i(t))H_i(t)  - 
 u_S H_i(t) (\avg{s_i}-s_0) \:,\\
  \frac{d P_{ij}(t)}{dt} & = &  \frac{1}{\tau} (1-\eta_i(t)) P_{ij}(t) - u P_{ij}(t)
  (\avg{s_i}-s_0)  - u_D D_{ij} P_{ij}(t) \:,
\end{eqnarray}
where $H_i\in [0,1]$ is now a probability of spontaneous firing,
$s_0$ is a target average activity and $D_{ij}$ is the distance between electrodes $i$ and $j$.
The input ratio is $\eta_i(t) = \sum_j P_{ij}(t)$.
Remember that, for a critical branching process,
$\avg{\eta_i} = 1$. 
These values were chosen as homeostatic targets. 

Shew \emph{et al.} \citep{Shew2015} studied
experimentally the visual cortex of the turtle
and proposed a (complete graph) 
self organizing model
for the input synapses $\Omega_i$ and the cortical
synapses $W_{ij}$. The stochastic
neurons fire with a linear saturating function:
\begin{eqnarray}
    \mathrm{Prob}(s_i[t+1] = 1) &=& \left\{\begin{array}{rcl}
V_i[t]   &\mathrm{if} & V < 1 \:, \\ & & \\
     1 &\mathrm{if} & V > 1\;,
    \end{array}\right.
    \\
    V_i[t] &=& \Omega_i[t] H_i[t] + \frac{1}{N} 
    \sum_j W_{ij}[t] s_j[t] \:,
\end{eqnarray}
where, like in Eq.~\eqref{eq:Hsu2006}, $H_i$ accounts for external stimuli. 
For both types of synapses they used the discrete
time LHG dynamics, Eq.~\eqref{discreteLHG},  
and concluded that the computational
model accounts very well for the experimental data.

Hernandez-Urbina and Herrmann~\citep{Hernandez2017}
studied a discrete time IF model where they define
a local measure called node success:
\begin{equation}
    \phi_j[t] = \frac{\sum_i A_{ij} s_i[t+1]}{\sum_i A_{ij}} \:,
\end{equation}
where $A$ is the adjacency matrix of the network, with $A_{ij}=1$ if $j$ projects onto $i$ ($A_{ij}=0$ otherwise). 
Note that we reversed the indices as compared with the original notation~\citep{Hernandez2017}.
Observe that $\phi_j$ measures how many postsynaptic neurons are excited by the presynaptic neuron $j$. 

The authors then define the node success driven
plasticity (NSDP):
\begin{equation}
W_{ij}[t+1]  =  W_{ij}[t] + 
\frac{1}{\tau} \exp\left(- \phi_j(t)/B\right)
- u \exp(-\Delta t_j/D) \:,
\end{equation}
where $\Delta t_j = t - \hat{t}_j$ is the 
time difference between the spike
of node $j$ occurring at current time step $t$ and its previous spike which occurred at $\hat{t}_j$ (the last spike), while $B$ and $D$ are constants.
Notice that the drive term is larger if the node success is small and the dissipation term is larger if the firing rate (inferred locally as $\hat{\rho} = 1/ \Delta t_j$) is large 
(compare with Eq.~\eqref{general}).

They analyzed the relation between the avalanche
critical exponents, the largest eigenvalue $\Lambda$
associated to the weight matrix and the data collapse
of the shape of avalanches for several network
topologies. All results are compatible with
(quasi-)criticality. They also found that 
if they stop NSDP and introduce STDP, the
criticality vanishes, but if the two
dynamics are done together, criticality
survives.

Levina \emph{et al.}~\citep{Levina2007b} proposed a model in a complete graph in which the branching ratio $\sigma$ is estimated as the local 
 branching $\sigma_i$ of
 a neuron that initiates an avalanche. 
 The homeostatic rule is to increase the synapses if $\sigma_i<1$ and decreasing them 
 if $\sigma_i>1$. 
 The network converges, with 
 SOqC oscillations, to $\sigma^* \approx \sigma_c = 1$.\\

\subsection{Meta-Plasticity}

Peng and Beggs~\citep{Peng2013} studied a square 
lattice ($K = 4$) of IF neurons with open 
boundary conditions.
A random neuron receives a small increment
 of voltage (slow drive). If the voltage of 
 presynaptic neuron
 $j$ is above a threshold $\theta = 1$,
 we have:
\begin{eqnarray}
V_j[t+1] &=& V_j[t] - 1 \:,\\
s_j[t+1] & = & \Theta(V_j[t+1]-\theta) \:,\\
V_i[t+1] &=& V_i[t] + \frac{1}{K} W_{ij}[t]s_j[t] \:,
\end{eqnarray}
The self-organization is made by a LHG dynamics plus a
meta-plasticity term:
\begin{eqnarray}
W_{ij}[t+1] & = & W_{ij}[t] + \frac{1}{\tau} \left( 
A - W_{ij}[t]\right) -  u W_{ij}[t] s_j[t] \:,\\
u_{a+1} &=& u_a - (1 - X_a)/N   \:,
\end{eqnarray}
where  $X_a$ is the total fraction of neurons at the boundary that fired during the $a$-th avalanche and $u_{a+1}$ is the updated value of $u$ after the avalanche. 
Notice that the meta-plasticity term differs from the MT model of Eq.~\eqref{MT}, because the hyperparameter $u$ is updated in a much slower time scale. 
Peng and Beggs show that this dynamics converges automatically to good values for the parameter $u$, that is, we no longer need set the $u$ value in advance.
We observe, however, that $X_a$ is a non-local variable.\\

\subsection{Hebbian Synapses} 

Ever since Donald Hebb's proposal that neurons that fire together wire together~\citep{Hebb1949,Turrigiano2000,Kuriscak2015}, several attempts have been made to implement this idea in models of self-organization. 
However, a pure Hebbian mechanism can lead to diverging synapses, so that some kind of normalization or decay needs also be included in Hebbian plasticity.

In 2006, de Arcangelis, Perrone-Capano and Herrmann introduced a neuronal network with Hebbian synaptic dynamics~\citep{Arcangelis2006} that we call the APH model.
There are several small variations in the models proposed by de Arcangelis \emph{et al.}, but perhaps the simplest one~\citep{Lombardi2017} is given by the following neuronal dynamics on a square lattice of $L\times L$ neurons:  
If at time $t$ a presynaptic neuron $j$ has a membrane potential above a firing threshold, $V_j[t] > \theta$, it fires, sending neurotransmitters to all its (non refractory) neighbors:
\begin{equation}\label{Arcangelis}
V_i[t+1] = V_i[t] + \overline{W}_{ij} V_j[t]\:,
\end{equation}
where $\overline{W}_{ij} = W_{ij}/\sum_l^{nn} W_{lj}$.
Then, neuron $j$ enters in a refractory
period of one time step.
The synaptic self-organizing dynamics is given by
\begin{eqnarray}
    W_{ij}[t+1] &=& W_{ij}[t] + \frac{1}{\theta}\:
    \overline{W}_{ij} V_j[t] \:,  \:\:\:\:\:\:\:\:\:\: (\mathrm{active\: synapses})\:,\\
    W_{ij} & \leftarrow & W_{ij} - \frac{1}{N_B} \sum_{ij}
    \delta W_{ij}\:, \:\:\:\:\:\:\:\:(\mathrm{ \: inactive\: synapses, \:after\: avalanche }) \:,\label{Arc}
\end{eqnarray}
where $N_B$ is the total number bonds and active 
(inactive) synapses are the ones used (not used) in
Eq.~\eqref{Arcangelis}. The sum in Eq.~\eqref{Arc} is 
over all synaptic modifications
$\delta W_{ij}[t+1] =  W_{ij}[t+1] - W_{ij}[t]$, a 
step which involves non-local information and amounts 
to a kind of synaptic rescaling. 
If the synaptic strength falls below some threshold, 
the synapse is deleted (pruning), so that this 
mechanism sculpts the network architecture.
So, co-activation of pre- and postsynaptic 
neurons makes the synapse grow, and inactive synapses
are depressed, which means that it is a Hebbian process. 
Several authors explored the APH model
in different contexts, including learning
phenomena~\citep{Pellegrini2007,Arcangelis2012,Arcangelis2012b,Lombardi2012,Arcangelis2014,Lombardi2014,Lombardi2014b,Kessenich2016,Lombardi2017}.

Çiftçi~\citep{Ciftci2018}  studied a neuronal SIRs model on the \emph{C. elegans} neuronal network topology.
The spontaneous activation rate (the drive) is $h = 1/\tau \rightarrow 0^+$  and the recovery rate to the susceptible state is $q$. 
The author studied the system as a function of $q/h$ (separation of time scales $q \gg h$).
The probability that a neuron $j$ activates its neighbor $i$ is
$P_{ij}$ ($g_{ij} = 1 - P_{ij}$ is the probability of synaptic failure in the author notation). 
The synaptic update occurs after an avalanche (of size $S$) and affects two neighbors that are active at the same time (Hebbian term): 
\begin{equation}
    P_{ij}[t+1] = \left \{ \begin{array}{cc}
 P_{ij}[t] + \frac{1}{\tau} \frac{1}{S} (1- P_{ij}[t])
&\mathrm{if\:the\:synapse\: was\:not\: used} \:,\\   \\
   P_{ij}[t] - u \left(1 - \frac{1}{S}
\right) P_{ij}[t] & \mathrm{if\:the\:synapse\:was\:used}
 \:. \end{array}
\right. 
\end{equation}
Ciftçi found robust self-organization to quasicriticality.
The author notes, however, that $S$ is a non-local information.

Uhlig \emph{et al.}~\citep{Uhlig2013} considered
the effect of  LHG synapses in the
presence of an associative Hebb synaptic matrix.
They found that, although the two processes are not 
irreconcilable,  the critical state 
has detrimental effects to the attrator
recovery. They interpret this as a suggestion
that the standard paradigm of memories as
fixed points attractors should be replaced by more 
general approaches like transient dynamics~\citep{Rabinovich2008}.

\bigskip
\subsection{Spike-Time Dependent Plasticity}

Rubinov~\emph{et al.}~\citep{Rubinov2011} studied
a hierarchical modular network of LIF
neurons with STDP plasticity. The synapses
are modeled by double exponentials:
\begin{eqnarray}
\frac{dV_i(t)}{dt} &=&  - (V_i(t)-E) + I + I_i^\mathrm{syn}(t) \:,\\
I_i^\mathrm{syn}(t) &  = & \sum_j W_{ij} V_0 \sum_{\hat{t}_j}
\left[\exp\left(- \frac{t - \hat{t}_j}{\tau_1}\right) - \exp\left(- \frac{t - \hat{t}_j}{\tau_2}\right) \right]  \:,
\end{eqnarray}
where $\{\hat{t}_j\}$ are the presynaptic firing times.
Synaptic weight changes at every spike of a presynaptic neuron, folowing the STDP rule:
\begin{eqnarray}\label{STDP}
    \Delta W_{ij}= \left\{ \begin{array}{ccc}
A_+(W_{ij}) \exp\left( -  \frac{\hat{t}_j - \hat{t}_i}{\tau_+}\right) & \mathrm{if} & \hat{t}_j 
< \hat{t}_i \:,\\ \\
- A_-(W_{ij}) \exp\left( -  \frac{\hat{t}_j - \hat{t}_i}{\tau_-}\right) & \mathrm{if} & \hat{t}_j 
\geq \hat{t}_i \:, 
    \end{array}\right. 
\end{eqnarray}
where $A_+(W_{ij})$ and $A_-(W_{ij})$ are weight dependent functions, see~\citep{Rubinov2011} for details. 
The authors show an association between modularity,
low cost of wiring, STDP and
self-organized criticality in a neurobiologically
realistic model of neuronal activity.

Del Papa \emph{et al.}~\citep{DelPapa2017} explored the interaction between criticality and learning in the context
of self-organized recurrent networks (SORN). The ratio between inhibitory to
excitatory neurons is $N^I/N^E = 0.2$. These neurons interact via
$W^{EE}, W^{IE}$ and $W^{EI}$ synapses (no 
inhibitory self-coupling). Synapses are
dynamic, and also the excitatory 
thresholds $\theta^E_i$. The neurons evolve as:
\begin{eqnarray}
 s^E_i[t+1] &=& \Theta \left( \sum_j^{N^E} 
 W_{ij}^{EE}[t] s^E_j[t] - \sum_k^{N^I} W^{EI}_{ik}
 s^I_j[t]  - \theta_i^E[t]  + I_i[t]
 + \eta_i^E[t] \right)\:,\\
 s^I_i[t+1] & = &  \Theta \left( \sum_j^{N^E} W^{IE}_{ij}
 s^E_j[t]  - \theta_i^I + \eta_i^I[t]
 \right) \:.
\end{eqnarray}
where $\eta_i[t]$ represents membrane noise.
Synapses and thresholds evolve following
five combined dynamics:
\begin{eqnarray}
W_{ij}^{EE}[t+1] & = & W_{ij}^{EE}[t] +
\frac{1}{\tau_\mathrm{STDP}} \left[ 
s^E_i[t+1] s^E_j[t] - s^E_j[t+1] s^E_i[t]\right] \:\:\:\:\: \mathrm{excitatory\: STDP} \:,\\
W_{ij}^{EI}[t+1] & = & W_{ij}^{EI}[t] -
\frac{1}{\tau_\mathrm{iSTDP}} s_j^I[t] \left[ 
1 -  s_i^E[t+1] (1+1/\mu_{\mathrm{IP}})
\right]
\:\:\:\:\: \mathrm{inhibitory\: STDP} \:,\\
W_{ij}[t+1] &\leftarrow& \frac{W_{ij}[t+1]}{\sum_j
W_{ij}[t+1]} \:\:\:\:\:\mathrm{
synaptic\: normalization\:(SN)} \:, \\
p(N^E) &=& \frac{N^E(N^E-1)}{N(N-1)} p(N)
\:\:\:\:\:
\mathrm{structural \: plasticity\:(SP)} \:,\\
\theta^E_i[t+1] & = & \theta^E_i[t]   +
\frac{1}{\tau_\mathrm{IP}} \left[ s^E_i[t]
- \mu_{\mathrm{IP}}
\right]\:\:\:\:\: \mathrm{intrinsic\: plasticity (IP)}\:,
\label{thetadyn}
\end{eqnarray}
where $\mu_{\mathrm{IP}} $ is the desired
activity level. In the structural plasticity process,
excitatory synapses are added with probability $p(N^E)$.
The authors found that this
SORN model presents well behaved power-law avalanche
statistics and that the plastic mechanisms
are necessary to drive the network to
criticality, but not to maintain it critical, that
is, the plasticity can be turned off after
the networks reaches the critical region.
Also, they found that noise was essential to
produce the avalanches, but degrade the 
learning performance. From this,
they conclude that the relation between
criticality and learning is more complex
and it is not obvious if criticality
optimizes learning.

Levina \emph{et al.}~\citep{Levina2014} studied the
combined effect of LHG synapses, 
homeostatic branching 
parameter $W_h$  and STDP:
\begin{equation}
    W_{ij}(t) = uJ_{ij}(t) W_h(t) W_\mathrm{STDP}(t) \:.
\end{equation}
They found that there is cooperativity of these mechanisms in extending the robustness of the critical state to variations on the hyperparameter $A$, see Eq.~\eqref{LHG}.

Stepp \emph{et al.}~\citep{Stepp2015} examined a LIF neuronal network which 
have both Markran-Tsodyks dynamics and spiking 
time-dependent plasticity STDP (both excitatory and
inhibitory). They found that, although 
MT dynamics produces
some self-organization, the STDP mechanism increses
the robustness of the network criticality.

Delattre \emph{et al.}~\citep{Delattre2015} included
in the STDP synaptic change $\Delta W_+$ 
a resource depletion term:
\begin{eqnarray}
\Delta W^\prime_+ &=& \gamma(\eta(t))\Delta W_+ \; ,\\
\gamma(\eta(t)) &=& \frac{1 - \exp\left(\frac{\eta^* - \eta(t)}{m}\right)}{1 + \exp\left(\frac{\eta^* - \eta(t)}{m}\right)} \; ,
\end{eqnarray}
where resource availability $\eta(t)$ evolves as:
\begin{equation}
    \frac{d\eta(t)}{dt} = \frac{1}{\tau_\eta} -
    \frac{\eta(t)}{\eta_0(\alpha(t)) \tau_\eta}\:.
\end{equation}
Here, $\alpha(t)$ is a continuous estimator of the network firing rate, $\tau_\eta$ is the recovery time of the resources availability and the term $\eta_0(\alpha(t))=(1+\alpha/k)^{-1}$ in the denominator  ensures that depletion is fast and recovery is slow ($k=20$~Hz).
They called this mechanism as network spiking dependent plasticity and showed that, in contrast to pure STDP, it leads to power-law avalanches with branching ratio around one.
\\
\subsection{Homeostatic Neurite Growth}

Kossio \emph{et al.}~\citep{Kossio2018}
studied IF neurons randomly distributed in a plane, with neurites distributed within circles of radii $R_i$ that evolved according to
\begin{equation}\label{R}
    \frac{dR_i}{dt} = \frac{1}{\tau} - u 
    \sum_{t_i} \delta(t-t_i) \:,
\end{equation}
where $\{t_i\}$ are the spike times of neuron $i$, with $\tau$ and $u$ constants.
Since the connections are given by
$W_{ij} = g O_{ij}$ where $g$ is a constant and $O_{ij}$ are the overlaping areas of the synaptic discs, Eq.~\eqref{R} is not much different from the simple synaptic dynamics of Eq.~\eqref{simpleLHG}, with constant drive and decay due to spikes.

Tetzlaff \emph{et al.}~\citep{Tetzlaff2010}
studied experimentally 
neuronal avalanches during the maturation of cell
cultures, finding that criticality is achieved in a
third stage of the dendrites/axons growth process.
They modeled the system using neurons with membrane potential
$V_i(t) < 1$ and calcium dynamics $C_i(t)$:
\begin{eqnarray}
\frac{dV_i(t)}{dt} &=& - \frac{V_i(t) - V_0}{\tau_V} 
+ \sum_j k_j^\pm W_{ij}(t) \:\Theta(V_j(t) - \eta_j(t)
)\:,\\
\frac{dC_i(t)}{dt}&=& -\frac{1}{\tau_C} C_i(t)
+ \beta \:\Theta(V_i(t) - \eta_i(t)) \:,
\end{eqnarray}
where $k^+ > 0$ $(k^- < 0)$  defines excitatory (inhibitory) neurons and $\eta_j(t) \in [0,1]$ is a random number.
Dendritic and axonal spatial distributions are again represented by their radii $R_i$ and $A_i$, whose dynamics are governed by calcium dynamics as:
\begin{eqnarray}
  \frac{dR_i(t)}{dt}  & =  &  - \frac{1}{\tau_R} (C_i(t) - 
  C_\mathrm{target}) \:,  \\
  \frac{dA_i(t)}{dt}  & =  &   \frac{1}{\tau_A} (C_i(t) - 
  C_\mathrm{target}) \:.
\end{eqnarray}
Finally, the effective connection is defined as:
\begin{eqnarray}
    W_{ij}(t)& =& \left[\gamma_1(t) - \frac{1}{2}\sin\left( 2 \gamma_1(t)\right)\right] A_j^2(t) +
    \left[\gamma_2(t) - \frac{1}{2}\sin\left( 2 \gamma_2(t)\right)\right] R_j^2(t) \:,\\
   \gamma_1(t) & = & \arccos\left( \frac{A_j^2(t) +
   D_{ij}^2 - R_i^2(t)}{2 A_j(t) D_{ij}}\right) \:,\:\:\:\:\:
     \gamma_2(t)  =  \arccos\left( \frac{R_i^2(t) +
   D_{ij}^2 - A_j^2(t)}{2 R_i(t) D_{ij}}\right)    \:,
\end{eqnarray}
where $D_{ij}$ is the distance between the neurons. 
This essentially represents the overlap of the axonal and dendritic zones, which can be understood as an abstract representation for the probability of synapse formation.

\section{Dynamic neuronal gains}

For all-to-all topologies as used
in~\citep{Levina2007,Levina2009,Bonachela2010,Brochini2016} the number of synapses is $N(N-1)$, which means that simulations become impractical for large $N$. 
Brochini \emph{et al.}~\citep{Brochini2016} discovered that, in their model with stochastic neurons, adaptation in a single parameter per neuron (the dynamic gain) is sufficient to self-organize the network. 
This reduces the number of dynamic equations from $\mathcal{O}(N^2)$ to $\mathcal{O}(N)$, enabling large-scale simulations.

The stochastic neuron has a probabilistic firing function, say,
a linear saturating function or a rational function:
\begin{eqnarray}
P(s = 1| V) = \Phi(V)  &=&  \Gamma (V-\theta) 
\:\Theta(V-\theta) \:\Theta(1-\Gamma (V-\theta)) 
+ \Theta(\Gamma (V -\theta) -1)\:, \\
P(s = 1| V) = \Phi(V)  &=& \frac{\Gamma (V-\theta)}{1 +\Gamma (V-\theta) }\:\Theta(V-\theta)
\:,
\end{eqnarray}
where $s= 1$ means a spike, $V$ is the membrane potential,
$\theta$ is the threshold
and $\Gamma$ is the neuronal gain.

Now, let's assume that each neuron $i$ has its neuronal gain $\Gamma_i$.
Several adaptive dynamics work, similar to LHG and even simpler:
\begin{eqnarray}
\Gamma_i(t+1) &=& \Gamma_i(t) +\frac{1}{\tau} \left[A - \Gamma_i(t) \right] -  u \Gamma_i(t) s_i(t) \:,
\label{G1}\\
\Gamma_i(t+1) &=& \Gamma_i(t) +\frac{1}{\tau} \Gamma_i(t)  -  u \Gamma_i(t) s_i(t) \:, \label{G2} \\
\Gamma_i(t+1) &=& \Gamma_i(t) +\frac{1}{\tau}  - u  s_i(t) \label{Gsimple}\:.
\end{eqnarray}
Costa \emph{et al.}~\citep{Costa2017} and Kinouchi \emph{et al.}~\citep{Kinouchi2019} studied the stability of the fixed points of mechanisms given by Eq.~\eqref{G1} and \eqref{G2} and concluded that the fixed point solution $(\rho^*,\Gamma^*)$ is of the form
$\rho^* = 0^+ + \mathcal{O}(1/\tau)$, 
$ \Gamma^* = \Gamma_c + \mathcal{O}(1/\tau)$. 
The fixed point is a barely stable focus for large $\tau$, which means that demographic noise  creates the hovering around the critical point (the sawtooth SOqC stochastic oscillations). 
The peaks of theses oscillations correspond to large excursions in the supercritical region, producing the so-called dragon king avalanches~\cite{Arcangelis2014}.

Zierenberg \emph{et al.}~\citep{Zierenberg2018}  considered a  cellular automaton neuronal model
with binary states $s_i$ and probabilistic synapses $P_{ij}[t] = \alpha_i[t] W_{ij}$, where $\alpha_i[t]$ is a homeostatic scaling factor.
The homeostasis is given by a negative feedback:
\begin{equation}\label{Pij}
    \alpha_i[t+1] = \alpha_i[t] + \frac{1}{\tau_\mathrm{hp}}
    \left(r^* - s_i[t]\right)
\end{equation}
where $\tau_\mathrm{hp}$ is the time constant of the homeostatic process 
and $r^*$ is a target level.
Notice that this mechanism depends only on the
activity of the postsynaptic neuron $i$, not the
presynaptic neuron $j$ as in the LHG model.
So, $\alpha_i[t]$ plays the same role of the 
neuronal gain $\Gamma_i[t]$ discussed above.

Indeed, for a cellular automata model similar to
~\citep{Costa2015,Campos2017}, a probabilistic synapse
with neuronal gains could be written as 
$P_{ij}[t] = \Gamma_i[t] W_{ij}$. In order to compare with the neuronal gain dynamics, we rewrite Eq.~\eqref{Pij} as:
\begin{equation}
    \Gamma_i[t+1] =  \Gamma_i[t] + \frac{1}{\tau}  
    - u  s_i[t]\:,
\end{equation}    
where $\tau = \tau_\mathrm{hp}/r^*$ and $u = 1/\tau_\mathrm{hp}$. So, in Zierenberg \emph{et al.},
we have a neuronal gain dynamics similar
to Eq.~\eqref{simpleLHG}, with hovering around the critical point and the ubiquitous sawtooth oscillations in $\alpha[t] \equiv \avg{\alpha_i[t]}$.

\section{Adaptive firing thresholds}\label{thres}

Girardi-Schappo \emph{et al.}~\citep{Girardi2020} examined a network with $N_E=pN=0.8 N$ excitatory and $N_I=qN=0.2 N$ inhibitory stochastic LIF neurons.
They found a phase diagram very similar to that of the Brunel model~\citep{Brunel2000}, with synchronous regular (SR), asynchronous regular (AR), synchronous irregular (SI) and asynchronous irregular (AI) states.
Close to the balanced state $g = W^{II}/W^{EE} = p/q = 4$
they found an absorbing-active second order phase transition with a critical point
$g_c = p/q - 1/(q \Gamma W^{EE})$. 
The self-organization of the $W^{II}$ and $W^{EI}$ inhibitory synapses was accomplished by a LHG dynamics.

They noticed, however, that for these stochastic LIF systems, the critical point requires also a zero field 
$h = I - (1-\mu) \theta$,
where $I$ is the external input and $\mu$ is the leakage parameter. 
While setting $h = 0$ for the critical point of spin systems is natural, obtaining zero field in this case demands self-organization, which is done by an adaptive firing threshold:
\begin{equation}\label{theta}
    \theta_i[t+1] = \theta_i[t] - \frac{1}{\tau_\theta} \theta_i[t]
    + u_\theta \theta_i[t] s_i[t] \:.
\end{equation}
Notice the plus signal in the last term, since if the postsynaptic neuron fires ($s_i = 1$) then the threshold must increase to hinder new firings. 
This mechanism is biologically plausible and also explains  classical firing rate adaptation.
Remembering that $\rho = \avg{s_i} \propto
h^{1/\delta_h}$ in the critical 
point, where $\delta_h$ is the field critical
exponent, from Eq.~\eqref{theta} we
have $h \propto 1/(\tau_\theta 
u_\theta)^{\delta_h} \approx 0$ for large $\tau_\theta u_\theta$.

As already seen, Del Pappa \emph{et al.}~\citep{DelPapa2017} considered a similar threshold dynamics, Eq.~\eqref{thetadyn}.
Bienenstock and Lehmann~\citep{Bienenstock1998} also studied, at the mean field level, the joint evolution of firing thresholds and dynamic synapses (see Sec.~\ref{Sync}).


\section{Topological Self-Organization}

Consider a cellular automata model~\citep{Kinouchi2006,Assis2008,Costa2015,Campos2017} in a  network with average degree $K$ and average probabilistic synaptic weights $P = \avg{P_{ij}}$. 
The critical branching ratio is $\sigma = P K = 1$, that is, 
critical average weight $P_c = 1/K$. Notice that we can
study networks with any $K$, even the complete graph,
where $P_c = 1/(N -1)$. In this network, what is
critical is the activity, which does not depends of the topology
(the degree $K$).

In another sense, we call a network topology critical if there is a barely infinite percolating cluster, which 
for a random network occurs for $K_c=2$. Several authors, starting in 2000 with Bornholdt and Rohlf~\citep{Bornholdt2000}, explored the self-organization toward this type of topological criticality~\citep{Bornholdt2003,Rohlf2008,Gross2009,
Rohlf2009,Meisel2009,Min2009,Hesse2014,Rybarsch2014,Cramer2020}.

So, we can have a critical network with a $W_c$ 
and any $K$ or a
topologically critical network with a well 
defined $K_c$.
The two concepts (activity criticality and topological criticality)
are different, but sometimes a topological criticality
also presents a phase transition 
with power-law avalanches and critical phenomena.
The topological phase transition
is  continuous and has a critical point, related to
the formation of a percolating cluster of nodes, but in the Bornholdt and Rohlf (BR) model it is related to an order-chaos phase transition, not 
to an absorbing state phase transition.

We  present here a more advanced version of the BR model~\citep{Bornholdt2003}. 
It follows the idea of deleting synapses from correlated
neurons and increasing synapses of uncorrelated neurons.
The correlation over a time $T$ is calculated as:
\begin{equation}
    C_{ij}[T] = \frac{1}{T+1} \sum_{t = t_0}^{t_0 + T}
    s_i[t] s_j[t] \:,
\end{equation}
where the stochastic neurons evolve as:
\begin{eqnarray}
V_i[t+1] &=& \sum_j W_{ij} s_j[t] \:, \\
\mathrm{Prob}\left(s_i[t+1] = + 1 \right) &=& \Phi(V_i) \:, \\
\mathrm{Prob}\left(s_i[t+1] = -1 \right) &=& 1 - \Phi(V_i) \\
\Phi[V_i] &=& \frac{1}{1 + \exp(-2 \Gamma (V_i - \theta_i))}
\end{eqnarray}
The self-organization procedure is:
\begin{itemize}
\item Choose at random a pair $(i,j)$ of neurons;
\item Calculate the correlation $C_{ij}(T)$;
\item Define a threshold $\alpha$.  
If $C_{ij}(T)>\alpha$, $i$
receives a new link $W_{ij}$ randomly drawn from a uniform distribution on $[-1,1]$ from site $j$, and if 
$C_{ij}<\alpha$ the link is deleted.
\item Then, continue updating the network state $\{s_i\}$
and self-organizing the network.
\end{itemize}

Interesting analytic results for this class of topological
models were obtained by Droste \emph{et al.}~\citep{Droste2013}.
The self-organized connectivity is about $K_c \approx 2$,
where the order-chaos transition occurs.
We must notice, however, that $K = 2$ seems to be
a very low degree for biological neuronal networks.
Kuehn~\citep{Kuehn2012} studied how the topological dynamics
time scale $\tau$ and noise level $D$ affects the BR
model, finding that optimal convergence to the critical point
occurs for finite values of $\tau_{\mathbf{opt}}$ and 
$D_\mathbf{opt}$.

Zeng \emph{et al.}~\citep{Zeng2015} combined the rewiring rules of the BR model with the neuronal dynamics of the APH model. 
They obtained an interesting result: the final topology is a small world network with a large number of neighbors, say $\avg{K} \approx 100$.
This avoids the criticism made above about the low number $K \approx 2$ of the BR model.


\section{Self-organization to other phase transitions}

\bigskip 
\subsection{First-order Transition}

Mejias \emph{et al.}~\citep{Mejias2010} studied a neuronal
population model with firing rate $\nu(t)$, which can be written in terms 
of the firing density $\rho = \nu/\nu_\mathrm{max}$:
\begin{equation}
   \tau_\rho \frac{d \rho}{dt} =  - \rho
    + S(W(t) \rho - \theta) + D_\eta 
    \eta(t) \:,
\end{equation}
where $S(z) = (1/2) [1+\tanh(z)]$ is a
(deterministic) firing function, $\eta(t)$ is a zero-mean Gaussian noise and $D_\eta$ is a noise amplitude. 
They used a depressing average synaptic weight inspired by a noisy LHG model:
\begin{equation}
    \frac{d W(t)}{dt} =  \frac{1}{\tau} \left[1-W(t) \right] - u W(t) \rho(t)
    + D_W \eta(t) \:.
\end{equation}
where $D_W$ is the synaptic noise amplitude.
Within a certain range of noise, they observed up-down states with irregular intervals, leading to a distribution of permanence times $T$ in the up state as $P(T) \propto T^{-3/2}$.
Notice that this model already starts with the mean-field equations, it is not a microscopic model (although a microscopic model perhaps could be constructed from it).

Millman \emph{et al.}~\citep{Millman2010} 
obtained similar results at a first order phase 
transition, but now in a random network of LIF neurons 
with average of $K$ neighbors and chemical synapses.
The synapses follow the LHG mechanism:
\begin{equation}
    \frac{dW_{ij}(t)}{dt} = \frac{1}{\tau}\left[A - W_{ij}(t) \right]
    - u W_{ij}(t) s_j(t) \:,
\end{equation}
where $W_{ij}(t) = p_r U_{ij}(t)$ in the 
authors notation ($p_r$ for
probability of releasing vesicles, $U_{ij}(t)$ for
synaptic resources) and $A = p_r$. They found that
the branching ratio is close to one in the 
up state, with power-law avalanches with size
exponent $3/2$ and lifetime exponent $2$.

Di Santo \emph{et al.}~\citep{diSanto2016,diSanto2018} 
and Buendia~\emph{et al.}~\citep{Buendia2020,Buendia2020b} studied
the self-organization toward a first-order phase transition
(called self-organized bistability or SOB). 
The simplest self-organizing dynamics was used in a two-dimensional model:
\begin{eqnarray}
    \frac{d\rho(\vec{x},t)}{dt} &=& \left[a +\omega E(\vec{x},t)\right]\rho(\vec{x},t) -
    b \rho^2(\vec{x},t) - \rho^3(\vec{x},t)  + D \nabla^2 \rho(\vec{x},t) +
    \eta(\vec{x},t) \:,\\
    \frac{dE(\vec{x},t)}{dt} &=& \nabla^2 \rho(\vec{x},t)+ \frac{1}{\tau}\left[A-E(\vec{x},t)\right] - u \rho(\vec{x},t)\:,
\end{eqnarray}
where $\omega, a> 0,b<0$ are constants, $A$ is the
maximum level of charging, $D$ is the diffusion constant
and $\eta(\vec{x},t)$ is a zero-mean Gaussian noise with
amplitude $\rho$. The authors original notation is 
$h = 1/\tau, \epsilon = u$ and 
$E$ is a (former) control parameter.
In the limit $1/\tau \rightarrow 0^+, u \rightarrow 0^+,
1/(\tau u) \rightarrow 0$,
this self-organization is conservative and can produce a tuning to the Maxwell point with power-law avalanches (with mean-field exponents) and dragon-king quasi-periodic events.  

Relaxing the conditions of infinite separation of time scales and bulk conservation, 
the authors studied the model with an LHG 
dynamics~\citep{diSanto2018,Buendia2020,Buendia2020b}:
\begin{eqnarray}
    \frac{d\rho(\vec{x},t)}{dt} &=& [a+W(\vec{x},t)]\rho(\vec{x},t) -
    b \rho^2(\vec{x},t) - \rho^3(\vec{x},t) + I  + D \nabla^2 \rho(\vec{x},t) +
    \eta(\vec{x},t) \:,\\
    \frac{dW(\vec{x},t)}{dt} &=&  \frac{1}{\tau}
    \left[A - W(\vec{x},t) \right] - u W(\vec{x},t) \rho(\vec{x},t) \:,
\end{eqnarray}
where $W$ is the  synaptic weight and $I$ an small input.
They found that this is the equivalent SOqC version
for first order phase transitions, obtaining histeretic
up-down activity, which has been called self-organized
collective oscillations
(SOCO)~\citep{diSanto2018,Buendia2020,Buendia2020b}.
They also observed bistability phenomena.

Cowan \emph{et al.}~\citep{Cowan2013} also found hysteresis cycles due to bistability in a IF model from the combination of an excitatory feed-back loop with anti-Hebbian synapses in its input pathway. 
This leads to avalanches both in the up state and in the down state, each one with power-law statistics (size exponents close to $3/2$). 
The hysteresis loop leads to a sawtooth oscillation in the average synaptic weight. 
This is similar to the SOCO scenario.

\bigskip

\subsection{Hopf bifurcation}

Absorbing-active phase transitions are associated to transcritical bifurcations in the low-dimensional mean-field description of the order parameter.  
Other bifurcations (say, between fixed points and periodic orbits) can also appear in the low-dimensional reduction of systems exhibiting other phase transitions, such as between steady states and collective oscillations. 
They are critical in the sense that
they present phenomena like critical slowing down (power-law relaxation to the stationary state), critical exponents etc.
Some authors explored the homeostatic self-organization toward such bifurcation lines.
 
In what can be considered a precursor in this field, Bienenstock and Lehmann~\citep{Bienenstock1998} proposed to apply a Hebbian-like dynamics at the level of rate dynamics to the Wilson-Cowan equations, having shown that the model self-organizes near a Hopf bifurcation to/from oscillatory dynamics. 
 
The model has excitatory and inhibitory stochastic neurons. 
The neuronal equations are:
\begin{eqnarray}\label{modelBL}
V_i^E(t) &=& \sum_j W_{ij}^{EE} s^E_j(t) + \sum_j W_{ij}^{EI} s^I_j(t) - \theta^E_i \:,\\
V_i^I(t) &=& \sum_j W_{ij}^{IE} s^E_j(t) + \sum_j W_{ij}^{EI} s^I_j(t) - \theta^I_i \:,
\end{eqnarray}
where, as before,  the binary variable $s \in \{0,1\}$ denotes the firing of the neuron. 
The update process is  
an asynchronous (Glauber) dynamics: 
\begin{equation}
P(s = 1 | V) = \frac{1}{2} \left[ 1+ \tanh (\Gamma  V(t)) \right] \: ,
\end{equation}
where $\Gamma$ is the neuronal gain.
 
The authors proposed a covariance-based regulation for the synapses $W^{EE}$ and $W^{IE}$ and a homeostatic process for the firing thresholds $\theta^E(t), \theta^I(t)$. 
The homeostatic mechanisms are:
\begin{eqnarray}
\frac{d W^{EE}(t)}{dt} & = & \frac{1}{\tau_{EE}} \left(
c^{EE}(t) - \Theta^{EE} \right) \:,\label{we}  \:\:\:\:\:\:\:\:\:\:
\frac{d W^{IE}(t)}{dt}  =  -\frac{1}{\tau_{IE}} 
\left(c^{IE} - \Theta^{IE} \right) \:, \\
\frac{d \theta^E(t)}{dt} & = & \frac{1}{\tau_{E}} 
\left(\rho^E(t) - \Theta^{E} \right) \:,  \:\:\:\:\:\:\:\:\:\:\:\:\:\:\:\:\:\:\:\:\:\:\:
\frac{d \theta^I(t)}{dt}  =  \frac{1}{\tau_{I}} 
\left(\rho^E(t) - \Theta^{I} \right) \:, \label{ti}
\end{eqnarray}
where $c^{EE} \equiv \left(\rho^E(t) - 
\avg{\rho^E(t)}\right)^2 $
is the variance of the excitatory activity $\rho^E(t)$,
$c^{IE}\equiv \left(\rho^E(t) - \avg{\rho^E}\right)
\left(\rho^I(t) - \avg{\rho^I}\right) $ is the excitatory-inhibitory covariance, $\tau_{EE},\tau_{IE},\tau_E, \tau_I$ are time constants, and $\Theta^{EE},\Theta^{IE},\Theta^{E},\Theta^{I} $ are target constants.
 
The authors show that there are Hopf and saddle-node lines in this system and that the regulated system self-organizes at the crossing of these lines. 
So, the system is very close to the oscillatory bifurcation, showing great sensibility to external inputs.
 
As commented, this paper is a pioneer in the sense of searching for homeostatic  self-organization at a phase transition in a neuronal network in 1998, well before the work of Beggs and  Plenz~\citep{Beggs2003}. 
However, we must recognize some deficiencies that later models tried to avoid. 
First, all the synapses and thresholds have the same value, instead of an individual dynamics for each one, as we saw in the preceding sections.
Most importantly, the network activities $\rho^E$ and $\rho^I$ are non-local quantities, not locally accessible to Eqs.~\eqref{we}-\eqref{ti}.
 
Magnasco \emph{et al.}~\citep{Magnasco2009} examined a very stylized model of neural activity with time dependent anti-Hebbian synapses:
\begin{eqnarray}
\frac{dV_i(t)}{dt} &=& \sum_j W_{ij}(t) V_j(t) \:,\\
\frac{dW_{ij}(t)}{dt} & = & \frac{1}{\tau} \left(\delta_{ij} - 
V_i(t) V_j(t) 
\right) \:.
\end{eqnarray}
where $\delta_{ij}$ is the Kronecker delta.
They found that the system self-organizes around a Hopf bifurcation, showing power-law avalanches and hovering  phenomena similar to SOqC.
 
\bigskip 
 
\subsection{Edge of synchronization}\label{Sync}

Khoshkhou and Montakhab~\citep{Khoshkhou2019} 
studied a random network with  $K = \avg{K_i}$ neighbors. 
The cells are Izhikevich neurons described by: 
\begin{eqnarray}
\frac{dV_i(t)}{dt} &=& 0.04 V_i^2(t) + 5 V_i(t) + 140 
- u_i(t)
+ I + I^{\mathrm{syn}}_i(t) \:,\\
\frac{du_i(t)}{dt} & = & a(b V_i(t) - u_i(t)) \:\\
\mathrm{if}\:\: V_i \geq 30 &\mathrm{then}& V_i \leftarrow c, 
u_i \leftarrow u_i + d \:.
\end{eqnarray} 
The parameters $a,b,c$ and $d$ are chosen to have regular spiking excitatory neurons and fast spiking inhibitory neurons. The synaptic input is composed of chemical double-exponential pulses with time constants $\tau_s$ and $\tau_f$:
\begin{equation}
    I^{\mathrm{syn}}_i = \frac{V_0 - V_i}{K_i (\tau_s - \tau_f) }
    \sum_j W_{ij} \left[\exp\left(- \frac{t-(t_j + 
    \tau_{ij}) }{\tau_s}\right)  
    - \exp\left(- \frac{t-(t_j + 
    \tau_{ij}) }{\tau_f}\right)
     \right]\:,
\end{equation}
where $\tau_{ij}$ are axonal delays from $j$ to $i$, 
$V_0$ is the reversal potential of the synapses, and $K_i$ is the in-degree of node $i$.

The inhibitory synapses are fixed but the excitatory
ones evolves with a STDP dynamics. If the firing
difference is $\Delta t = t_\mathbf{post} - 
t_\mathbf{pre}$, when the postsynaptic neuron $i$
fires, the synapses change by:
\begin{equation}
    \Delta W_{ij} = \left\{\begin{array}{ccc}
        A_+ \left(W_{\mbox{max}} - W_{ij}\right) 
        \exp\left(-\frac{\Delta t - \tau_{ij}}{\tau_+}\right) & \mathbf{if}
        & \Delta t > \tau_{ij} \:,\\
    A_- \left(  W_{ij} - W_{\mbox{min}}\right) 
        \exp\left(-\frac{\Delta t - \tau_{ij}}{\tau_-}\right)    &\mathbf{if}   & \Delta t \leq \tau_{ij} \:.
    \end{array}        \right.
\end{equation}
This system presents a transition from 
out-of-phase to synchronized spiking.
The authors show that a STDP dynamics self-organizes 
in a robust way the system to the border of this
transition, where critical features like 
avalanches (coexisting with oscillations) appear.

\section{Concluding Remarks}

In this review we described several examples of self-organization mechanisms that drive neuronal networks to the border of a phase transition (mostly a second order absorbing phase transition, but also to first order, synchronization, Hopf and order-chaos transitions).
Surprisingly, for all cases, it is possible to detect neuronal avalanches with mean-field exponents similar to those obtained in the experiments
of Beggs and Plenz~\citep{Beggs2003}.

By using a standardized notation, we recognized several
common features between the proposed homeostatic
mechanisms. 
Most of them are variants of the fundamental drive-dissipation dynamics of SOC and SOqC and can be grouped into a few classes.

Following Hernandez-Urbina and Herrmann~\citep{Hernandez2017}, 
we stress that the coarse tuning on hyperparameters of
homeostatic SOqC is not equivalent to
the fine tuning of the original control parameter.
This homeostasis is a \emph{bona-fide} self-organization, in the same sense that the regulation of body temperature is self-organized (although presumably there are hyperparameters in that regulation).
The advantage of these explicit homeostatic mechanisms is that they are biologically inspired and could be studied in future experiments to determine which are more relevant to cortical activity.

Due to non-conservative dynamics and the lack of an infinite separation of time scales, all these
mechanisms lead to SOqC~\cite{Buendia2020b,Bonachela2009,Bonachela2010}, not SOC.
In particular, conservative sandpile models should not be used to model neuronal avalanches because
neurons are not conservative.
The presence of SOqC is revealed by stochastic 
sawtooth oscillations in the former
control parameter, leading to large excursions in the
supercritical and subcritical phases.
However, hovering around the critical point seems to
be sufficient to account for the current experimental
data.
Also, perhaps the omnipresent  stochastic 
oscillations could be detected experimentally
(some authors conjecture that they are the 
basis for brain rhythms~\citep{Costa2017}). 

One suggestion for further research is to eliminate non-local variables in the homeostatic mechanisms.
Another is to study how the branching ratio $\sigma$, or better, the synaptic matrix largest eigenvalue $\Lambda$, depends on the self-organization hyperparameters (as done in Ref.~\cite{Campos2017}).
As several results in this review have shown, the dependence of criticality on the hyperparameters is always weaker than the dependence on the original control parameter.
Finally, one could devise local metaplasticity rules for the hyperparameters, similarly to
Peng and Beggs~\cite{Peng2013} (which, however, is unfortunatelly nonlocal).
An intuitive possibility is that, at each level of metaplasticity, the need for coarse tuning of hyperparameters decreases and criticality will turn out more robust.

\section*{Conflict of Interest Statement}

The authors declare that the research was conducted in the absence of any commercial or financial relationships that could be construed as a potential conflict of interest.

\section*{Author Contributions}

OK and MC contributed conception and design of the study; RP organized the database of revised articles; OK and MC
wrote  the manuscript. All authors contributed to manuscript revision, read and approved the submitted version.

\section*{Funding}
This article was produced as part of the activities of
FAPESP Research, Innovation and Dissemination Center
for Neuromathematics (Grant No. 2013/07699-0, S\~ao Paulo
Research Foundation). We acknowledge the financial support
from CNPq (Grants No. 425329/2018-6, 301744/2018-1 and
2018/20277-0), FACEPE (Grant No. APQ-0642-1.05/18), and Center
for Natural and Artificial Information Processing Systems (CNAIPS)-USP. 
Support from CAPES (Grants No. 88882.378804/2019-01 and 88882.347522/2010-01) and FAPESP (Grants No. 2018/20277-0 
and 2019/12746-3) is also gratefully acknowledged.

\section*{Acknowledgments}
OK thanks Miguel Muñoz for discussions and advice.




\bibliographystyle{frontiersinHLTH_FPHY} 
\bibliography{Bib}





\end{document}